\def\year{2021}\relax
\documentclass[letterpaper]{article} 
\usepackage{aaai21}  
\usepackage{times}  
\usepackage{helvet} 
\usepackage{courier}  
\usepackage[hyphens]{url}  
\usepackage{graphicx} 
\usepackage{natbib}  
\usepackage{caption} 
\usepackage{xcolor}
\usepackage{subfig}

\title{Cybersecurity Misinformation Detection on Social Media:  Case Studies on Phishing Reports and Zoom's Threats}

 \author{
    Mohit Singhal, Nihal Kumarswamy, Shreyasi Kinhekar, Shirin Nilizadeh \\
 }
 \affiliations{
    The University of Texas at Arlington\\
    (mohit.singhal, nihal.kumarswamy, shreyasi.kinhekar)@mavs.uta.edu, shirin.nilizadeh@uta.edu
 }

\begin{document}

\maketitle

\begin{abstract}

Prior work has extensively studied misinformation related to news, politics, and health, however, misinformation can also be about technological topics. While less controversial, such misinformation can severely impact companies' reputations and revenues, and users' online experiences. Recently, social media has also been increasingly used as a novel source of knowledgebase for extracting timely and relevant security threats, which are fed to the threat intelligence systems for better performance. However, with possible campaigns spreading false security threats, these systems can become vulnerable to poisoning attacks. In this work, we proposed novel approaches for detecting  misinformation about cybersecurity and privacy threats on social media, focusing on two topics with different types of misinformation: \emph{phishing websites} and \emph{Zoom's security \& privacy threats}. We developed a framework for detecting inaccurate phishing claims on Twitter. Using this framework, we could label about 9\% of URLs and 22\% of phishing reports as misinformation. We also proposed another framework for detecting misinformation related to Zoom's security and privacy threats on multiple platforms. Our classifiers showed great performance with more than 98\% accuracy. Employing these classifiers on the posts from Facebook, Instagram, Reddit, and Twitter, we found respectively that about 18\%, 3\%, 4\%, and 3\% of posts were misinformation. In addition, we studied the characteristics of misinformation posts, their authors, and their timelines, which helped us identify campaigns. 

\end{abstract}
\section{Introduction} 
Prior work has extensively studied misinformation related to news, politics, and health~\cite{rashkin2017truth,ruchansky2017csi,love2020parallel}. 
Though misinformation can also be about technologies and tools that people use in their everyday life. While less controversial, such misinformation can severely impact companies' reputations and revenues, and users' online experiences.
Moreover, recently social media has been increasingly employed as a novel source of information for threat intelligence systems.  
For example, a recent study showed that 25\% of vulnerabilities appear on social media before the National Vulnerability Database~\cite{PNNL}, and as the result, numerous threat intelligence tools, such as Spider-Foot~\cite{spiderfoot:2021} and IntelMQ~\cite{IntelMQ:2021}, started collecting  intelligence from social media platforms. In another recent study, it was shown that Twitter phishing reports provide detailed information about phishing threats, and include more sophisticated phishing threats that remain undetected by anti-phishing tools~\cite{roy2021evaluating}.  
However, we argue that such tools are vulnerable to poisoning attacks, because social media posts (1) can be posted by an adversary, and (2) are weakly monitored, as detecting and removing such misinformation is currently not the priority of social media platforms. 

On the other hand, to detect misinformation about technologies, novel approaches are needed to be employed, as compared to misinformation about news, politics, and health, their purpose and impact are different. One have roots in people's political and cultural views and beliefs, while the other might take advantage of people's lack of technical background, their beliefs about a technology or company, or their fear about possible security threats.  

No work has systematically studied the spread and characteristics of misinformation about technological topics. In this work, we study misinformation about cybersecurity and privacy threats on social media, focusing on: \emph{phishing websites} and \emph{Zoom's security \& privacy threats}. We chose these two because the misinformation about each of them has different intentions, characteristics, and consequences. By this work then we show that different approaches should be employed to detect these different types of misinformation. 

Phishing is a type of social engineering attack through which attackers try to trick victims into disclosing their private and sensitive information~\cite{dhamija2006phishing}. 
To inform other users, some social media users share reports of phishing websites~\cite{roy2021evaluating}. 
However, false phishing reports can also circulate on these platforms. 
These false reports can decrease the websites' visits, especially if these websites are added to blocklists. 
They can also increase the false-positive rates of anti-phishing systems. 
A recent work showed the presence of false phishing reports~\cite{roy2021evaluating} on Twitter, however, in this work, we systematically analyze this threat, and propose a detection algorithm that can with high accuracy identify false phishing reports in real-time.    

\begin{figure}[h]
\centering
   \includegraphics[width=0.65\linewidth]{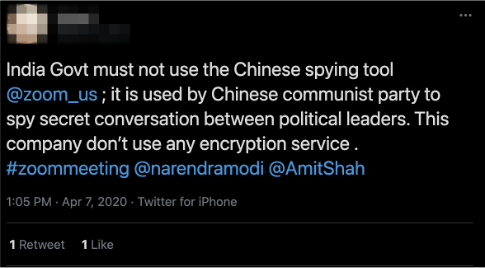}
   \caption{Claim about Zoom}
   \label{fig:figure-zoom-t}
\label{fig:example}
\end{figure}
With the surge in the use of video conferencing tools, such as Zoom~\cite{nyt:zoom} during the pandemic, came the concerns about the company handling of security and privacy of its user base. 
Users discussed issues, such as Zoombombing~\cite{zoombombingex}, and private Zoom meetings that can be available to the public~\cite{zoombombingex2}. 
However, not all the discussions were accurate. 
For example, Figure~\ref{fig:figure-zoom-t} shows a claim that Zoom is a ``Chinese spying tool.'' However, the author has not provided any supporting evidence, and the claim is misleading~\cite{politi}. In addition, the tweet claims that Zoom does not use any encryption service, which is not true on the date this tweet was posted~\cite{zoomwir}. 
Discourses like this can impact the overall image of the company, as well as Zoom's (new) users experiences. 
No other work has studied misinformation about Zoom, or any other technology related topic. Existing studies on fake news and vaccine misinformation also only focus on one social media platform, while we have investigated four platforms, including Twitter, Facebook, Instagram, and Reddit. 

Both phishing and Zoom misinformation can be categorized as \emph{fact-based} misinformation, as they are entities that have unique groundtruth values~\cite{kumar2018false}. For example, one can check if a website is phishing, or Zoom has a specific vulnerability. 
Therefore, we define misinformation based on criteria that help check these entities against their groundtruth values. 
We define a claim about a phishing website as a false report if the phishing link provided in the tweet refers to a benign website, and therefore the report is \textit{false}. 
We define a post about Zoom's security and privacy threats as misinformation if the post fails to provide any supporting evidence and cannot be verified by cross-checking it with reputable sources, or the provided information is fully or partially in contrast with that of trusted sources. While both false phishing reports and Zoom claims can be intentional, and therefore disinformation, in this paper, we do not differentiate between misinformation and disinformation, as we do not explore methods to detect intentions. 

The examples and definitions of misinformation for these topics illustrate that not the same approach can be employed to detect these different types of misinformation. To detect a false phishing report, one naive solution is to check if they appear on open anti-phishing sites, such as PhishTank, or VirusTotal. However, these sites might not be up-to-date~\cite{peng2018detecting,kantchelian2015better}. To detect misinformation about Zoom security and privacy, not only does one need to have access to a knowledgebase that lists security and privacy threats of Zoom but one also need to consider the language, used by the author, to discuss the matter, as, unlike the phishing reports, the discussion over Zoom does not follow a certain structure/ template and users might be addressing the issue arbitrarily. 

In this paper, we investigate three research questions: \textbf{RQ1:}~How misinformation about phishing and Zoom are prevalent on social media? \textbf{RQ2:}~How such misinformation can be detected?
\textbf{RQ3:}~What are the characteristics of misinformation posts, their authors, and campaigns?

To answer these research questions, we propose two different frameworks for detecting and analyzing misinformation related to each of these topics. 
The first framework investigates the correctness of phishing reports on Twitter through a multi-step approach, which includes (1) obtaining tweets about phishing, extracting unique URLs, (2) creating a groundtruth by regularly checking them via VirusTotal~\cite{VirusTotalAPI:2020}, on PhishTank~\cite{phishtank} and manually, and then (3) developing a classifier that can successfully detect false phishing reports in real-time. 
The second framework examines the correctness of posts about \emph{Zoom}'s security and privacy threats by (1) obtaining posts from Facebook, Instagram, Reddit, and Twitter, (2) using a ground truth dataset to identify the features that make misinformation posts distinguishable from accurate posts, and (3) using the features to build a classifier that detects misinformation. 
Thus, in this paper, we have the following contributions: 
\begin{enumerate}
\item We showed social media users share false phishing reports and proposed a framework for detecting such posts. 
\item We presented a new annotated groundtruth dataset for security \& privacy issues regarding Zoom, and identified misinformation features through quantitative and qualitative analysis. This dataset can be used as a benchmark by the community to build and test their proposed detection algorithms. 

\item We developed classifiers that detect misinformation about Zoom's security and privacy on four different social media platforms. 
Using these classifiers, we showed that such misinformation is prevalent, especially on some platforms, such as Facebook. 

\item We characterized the detected misinformation posts, their authors, and possible campaigns.

\item We hope that this work increases the awareness of the community and social media platforms about the spread of misinformation about technological topics.

\end{enumerate}

\section{Related Work}
\textbf{Obtaining cybersecurity threats from social media.} 
Recently, some works have proposed using social media, such as Twitter and Facebook, as the main source of identify new vulnerabilities~\cite{191006, roy2021evaluating}.
Alves et al.~\cite{alves2021processing} introduced a Twitter streaming threat monitor that generates a continuously updated summary of the threat landscape related to a monitored infrastructure. 
Okutan et al.~\cite{okutan2017predicting} integrated tweets with posts from the GDELT news service and Hackmageddon to detect new threats related to one of three topics: Defacement, DoS, and Malicious Email/URL. 
Sapienza et al.~\cite{sapienza2017early} introduced a system that leverages the communication of malicious actors on the dark web, and the activity of cyber security experts on Twitter to automatically generate warnings of imminent or current cyber-threats. 

\textbf{Misinformation Detection in Social Media}. 
A large body of work has tried detecting fake political news~\cite{shu2017fake,shu2019beyond,zhou2020safe}, investigating various linguistic features~\cite{hosseinimotlagh2018unsupervised,markowitz2014linguistic}, as well as deep neural networks~\cite{karimi2018multi,karimi2019learning,wang2018eann}. 
In recent months, scholars have analyzed misinformation related to COVID-19~\cite{brennen2020types,kouzy2020coronavirus,loomba2021measuring,singh2020first}. A recent work~\cite{roy2021evaluating} showed the presence of false phishing reports on Twitter. We analyze this threat in more depth and propose an algorithm for detecting false phishing reports. To the best of our knowledge, our work is the first study to find misinformation about security claims.

\section{Framework for Detecting False Phishing Reports on Twitter}
\begin{figure*}[ht]
    \centering
    \includegraphics[width=0.9\textwidth]{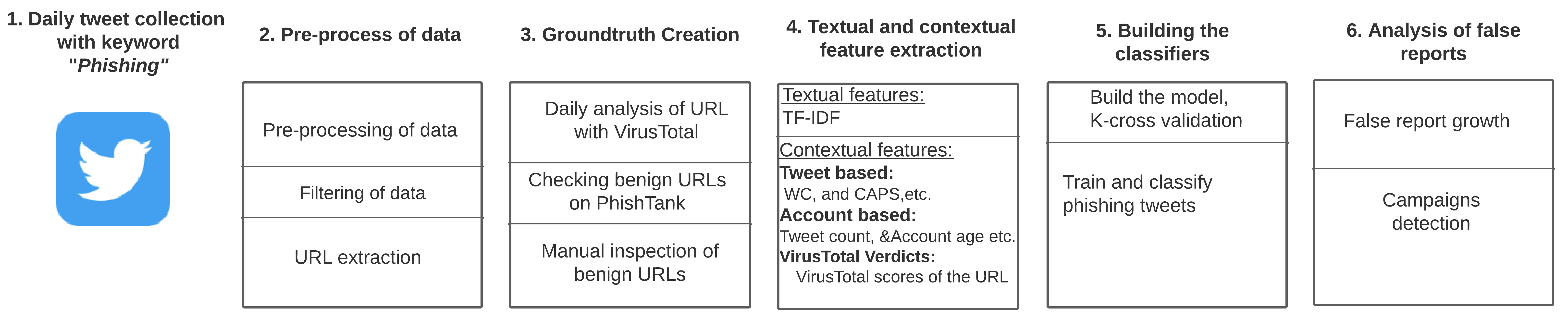}
    \caption{The framework developed for detecting false phishing reports on Twitter}
    \label{fig:phishing-datasetcollect}
\end{figure*}

We propose a framework for detecting false phishing reports on Twitter. 
We conduct our study on Twitter because other works have shown the existence of phishing reports on this platform~\cite{roy2021evaluating}. 
 Figure~\ref{fig:phishing-datasetcollect} shows our proposed framework, which consists of (1) Data collection, (2) Groundtruth creation, (3) Feature selection, (4) Classification, and (5) Analysis of false phishing reports.

\subsection{Data Collection} 
The data collection module is about collecting phishing reports on Twitter, and it consists of three steps: 
\noindent\textbf{(1) Collection of tweets:} Using the Twitter streaming API~\cite{Twitter:2020}, we collected 1\% sample of daily tweets on Twitter that include the keyword \emph{phishing}, from January 11, 2021, to April 11, 2021. We did not use Twitter V2 API, because at the time of data collection it was not available~\cite{Twitter-v2faq}. We considered retweets in our dataset.
\textbf{(2) Pre-processing and filtering:} We filtered tweets to obtain those that were posted on the same day, as the Twitter API provides tweets posted over the last 7 days. 
\textbf{(3) URL extraction:} After manually inspecting a random sample of 100 tweets, we found that many tweets are \emph{educational}, and are about phishing threats in general, or provide some stories about phishing events. 
We also found that the tweets with phishing reports usually include the obfuscated URL of the potential phishing website, i.e., using the following formats ``hxxps[:]//xyz[.]com,'' or ``hXXp:[//]xyz[dot]com.'' This is consistent with observation provided by~\cite{roy2021evaluating}.  
Therefore, we employed regular expressions to retrieve the tweets with obfuscated URLs and obtain those tweets claiming about specific websites being phishing websites.  
We manually checked the obtained 100 tweets and found that the precision of this module is 1, and our final dataset only includes tweets with phishing claims. Finally, we extracted the URLs from these tweets to be validated. 

\subsection{Groundtruth Creation} 
For training the classifier, we created a groundtruth dataset of true and false phishing reports. This is not a trivial labeling task as it required a longitudinal analysis of phishing reports. This module consists of the following steps:   
\textbf{(1) Daily analysis of URLs via VirusTotal: } 
We evaluated the URLs \emph{daily} by passing them through VirusTotal API~\cite{VirusTotalAPI:2020}. 
VirusTotal provides aggregated results for an URL obtained from 80 scan engines by third-party security vendors. Given a particular URL, the API returns the labels from all the vendors, and it shows the number of scan engines that detected the URL as \emph{malicious} and \emph{benign}. 
We analyzed the URLs daily because phishing campaigns can be short-lived, and some websites can go inactive, or be re-sold. 
\textbf{(2) Delayed analysis of VirusTotal scores: } 
Prior work~\cite{vtpaper_blackbox,kantchelian2015better,zhu2020measuring} has pointed out that VirusTotal is slow in updating its database, and there is a high probability that a website that was flagged as benign on the first day, be flagged malicious later on. 
Therefore, we employed VirusTotal again on May 3, 2021, three weeks after the last day of data collection. 
\textbf{(3) Checking benign URLs on PhishTank:} Since anti-phishing engines can misclassify malicious URLs as benign, this module further passed all the URLs labeled as \emph{benign} to PhishTank~\cite{phishtank}, a free community-based site for reporting phishing websites.
\textbf{(4) Manual inspection of \emph{benign} URLs:} Since it is also possible that the malicious URLs are not detected by VirusTotal and also have not been reported to the PhishTank, additionally, we manually checked the URLs that are labeled \emph{benign} in the previous steps. 
Particularly, on a virtual machine, we checked whether the URL is mimicking a login page or prompted users to download something. 

\begin{table*}[t]
  \caption{We found 22\% of phishing reports to be false, as they were inaccurately reported as phishing websites.}
  \centering
  \label{tab:freq4}
\resizebox{0.8\textwidth}{!}{%
  \begin{tabular}{p{0.05\textwidth}p{0.05\textwidth}p{0.07\textwidth}p{0.07\textwidth}p{0.08\textwidth}p{0.07\textwidth}p{0.13\textwidth}p{0.12\textwidth}p{0.12\textwidth}}
  \hline 
  No. of Tweets & Unique Users & Unique URLs & Malicious URLs & Benign URLs & Malicious Tweets & Benign Tweets (False claim) & Accounts with true claims & Accounts with false claims\\
  \hline
     17,770 & 11,472 & 10,578 & 9,603 & 975 (9\%) & 13,875 & 3,895 (22\%) & 11,200 & 148 \\
    \hline
\end{tabular}
 }
\end{table*}

\textbf{Groundtruth dataset and prevalence of false phishing reports:} Table~\ref{tab:freq4} shows our final groundtruth dataset. 
We found a union set of 11,472 users, who posted 17,770 phishing reports. 
Among 10,578 unique URLs, we identified 9,603 as \emph{malicious} and 975 as \emph{benign}, corresponding to 13,875 and 3,895 tweets, respectively. 
Therefore, we can conclude that about 9\% of all \emph{obfuscated URLs}, and about 22\% of tweets are \emph{misinformation}. 
We found that 11,472 and 148 unique users with true and false reports, respectively, with 124 users posting a mix of true and false reports. 

\subsection{Detecting False Phishing Report in Real-time} 
To identify phishing misinformation in real-time, we developed a classifier using our groundtruth dataset. 

\subsubsection{Textual and Contextual Feature Selection} 
We used a combination of textual and contextual features. 
\textbf{Textual Features:} 
We extracted bi-grams and uni-grams from all the posts and considered the top 100 of them with the highest values of TF-IDF, which resulted in 64 uni-grams and 36 bi-grams. We tested our classifiers with 100, 500, 1000, 1500, and 2000 top n-grams, and found that, 100 provided the best results.
\textbf{Contextual Features:} Along with the textual features obtained from the posts, we extracted a set of contextual features from the meta-data, which include: (1)~\emph{VirusTotal Scores:} 
A Higher VirusTotal score can be a great indication of a URL being malicious. However, in line with prior work~\cite{vtpaper_blackbox,zhu2020measuring}, we found that VirusTotal scores might not be accurate, at the time of the report. Therefore, they cannot solely be trusted with detecting false phishing reports in real-time. We used VirusTotal scores, obtained at the time of Twitter data collection, as a feature in our classifier, and we also relied on additional features obtained from Twitter posts and authors, which might help distinguish a false report from a truthful report. 
(2)~\emph{Post-inspired features:} We used some features obtained from the posts, including \emph{length of tweet}, and \emph{the total number of capital words in a tweet}, which was used as a feature named \emph{CAPS}. (3) \emph{Features based on account characteristics:} 
Most prior work on phishing detection~\cite{sun2016automating,ma2009beyond,canali2011prophiler} use the features captured from the URLs. 
Some works~\cite{aggarwal2012phishari,chen20156}, on detecting phishing URLs on social media, have also used features extracted from the social media posts, such as author's features.   
We used the following account characteristics: \emph{number of tweets}, \emph{profile description length}, \emph{account age}, \emph{listed count}, \emph{verified account}, \emph{followers count}, and \emph{has a profile image}. These features indicate if accounts are well-established, active, and anonymous, which  might imply their trustworthiness~\cite{morris2012tweeting,zubiaga2014tweet}.

\subsubsection{Classifier} 
\label{phishing-classifier}
We developed a binary classifier, using our already labeled groundtruth dataset. We used the first 10 weeks of our groundtruth dataset for training and the remaining 3 weeks for testing. This setup evaluates the performance of our classifier on real-time data, i.e., phishing reports that are seen for the first time. 
We used 10,851 true claims posts, 2,434 false claims posts as our training set, and 3,024 \& 1,461 true and false claim posts for our testing set, respectively.
Before extracting the features, we performed pre-processing on our dataset, removed stop words, emojis, special characters (hashtags), and URLs and performed stemming, i.e., cataloging related words together. These steps help minimize the sparsity of data~\cite{da2014tweet,patwa2021fighting}, e.g., by not removing the \# sign from \#covid, the word \emph{covid} would have been counted separate from every other covid word present in the posts. We vectorized our data using the TF-IDF. We found that a mix of uni-gram \& bi-gram features provided a better result compared to just using either uni-gram or bi-gram or tri-gram. Since our groundtruth dataset for each platform was unbalanced, we employed several oversampling techniques, such as RandomOversampler, Synthetic Minority Over-sampling Technique (SMOTE)~\cite{chawla2002smote}, and Adaptive Synthetic Sampling (ADASYN)~\cite{he2008adasyn}. 
We also tested multiple classification algorithms, such as Random Forests, SVM, Naive-Bayes \& K-nearest neighbor. 

\begin{table}[h]
\centering
\caption{The performance of classifiers} 
\label{table:completepredict11}
\centering
    \resizebox{0.9\columnwidth}{!}{%
    \begin{tabular}{l| cccc}
    \hline 
    Classifier & Accuracy & F1 Score & Precision & Recall\\
    \hline 
    Only VirusTotal & 0.79 & 0.70  & 0.84 & 0.79 \\
       &  (+/- 0.01)&  (+/- 0.02) & (+/- 0.02)& (+/- 0.02)\\
    Only Contextual & 0.92  &0.93  & 0.93 & 0.92  \\
    & (+/- 0.02) & (+/- 0.01) & (+/- 0.03) &  (+/- 0.02) \\
    \textbf{All features } & \textbf{0.95 } & \textbf{0.95} & \textbf{0.95 } & \textbf{0.95 } \\
   & \textbf{ (+/- 0.02)} & \textbf{(+/- 0.03)} & \textbf{(+/- 0.02)} & \textbf{ (+/- 0.01)} \\
    \hline 
\end{tabular}
}
\end{table}

\subsubsection{Classifier Performance} 
Table~\ref{table:completepredict11} shows the results of our classifier. We found that only using VirusTotal scores, as the only feature, did not yield the best results. Adding other contextual features increased the performance, from 79\% accuracy to 92\%. We obtained the best results, i.e., 95\% accuracy, precision, recall, and F1-score when we used all the textual and contextual. 
We found that the Random Forest classifier and SMOTE provided the best performance. 
We also examined feature importance in the trained Random Forest model to understand which of the features have a higher importance in classification tasks. The top 5 features and their scores are: \emph{VirusTotal Score} (0.416), \emph{Length of Tweet} (0.086), \emph{Profile description length} (0.040), \emph{phishing} (0.032), \emph{possible threat} as a phrase (0.027). 

\subsection{Accounts and Campaign Characterization}
\begin{table}
\caption{Descriptive statistics of our final datasets.}
\centering 
\resizebox{\columnwidth}{!}{%
\begin{tabular}{lcccc|cccc}
\hline \\[-1.8ex] 
& \multicolumn{4}{c}{Users with true claims} & \multicolumn{4}{c}{Users with false claims} \\
\\[-1.8ex] 
Feature & \multicolumn{1}{c}{Mean} & \multicolumn{1}{c}{Min} & \multicolumn{1}{c}{Max} & \multicolumn{1}{c}{Median}  & \multicolumn{1}{c}{Mean} & \multicolumn{1}{c}{Min} & \multicolumn{1}{c}{Max} & \multicolumn{1}{c}{Median} \\  
\hline \\[-1.8ex] 
Followers & 6,210 & 0 & $\sim12M$ & 302& 2,049 & 0 & $\sim35K$ & 426 \\ 
Friends &   1,347 & 0 & $\sim275K$ & 416& 1,264  & 0 & $\sim22K$ & 458\\
Tweets &   29,972 &  1 & $\sim3.4M$ & 4,883& 60,293 & 6 & $\sim2.2M$ & 5,426\\ 
Verified & 0.02 & 0 & 1 & NA & 0.01 & 0 & 1 & NA\\
\hline 
\end{tabular} 
}
 \label{stats-freq5} 
\end{table} 

\subsubsection{Descriptive Statistics} 
Table~\ref{stats-freq5} statistically describes the user accounts with true and false claims. 
If a user had posted both true and false claims, we considered them in both sets. 
We compared the account characteristics of users with true claims vs. false claims. To compare features, such as \emph{Followers, Friends, \& Tweets}, we ran Mann-Whitney U tests as they do not follow a normal distribution. We could not reject the null hypothesis that users with \emph{false} and \emph{true} claims have the same distribution of followers counts, friends count, and tweet counts. We ran chi-square test for the \emph{verified} variable, and could not reject the null hypothesis. 

\subsubsection{Spread of Misinformation: Campaigns} 
We constructed a network based on \emph{following} relationships between all users in our false claim dataset using the Louvain Community Detection method~\cite{blondel2008fast}. In total, our network has 121 nodes and 618 edges. This is only a subset of the users, as we were not able to obtain the following 27 accounts. The average weighted degree is 5.1. The average path length from one randomly selected node to another is 2.47.  
In total, we were able to obtain 6  communities with 37, 34, 22, 22, 4, and 2 nodes. 
We found that 66 accounts posted the same false claims in their community. Also, the biggest community contributes to about 51\% of the false reports.

\begin{figure}[t]
    \centering
        \subfloat[Tweet counts]{\includegraphics[width=0.49\columnwidth]{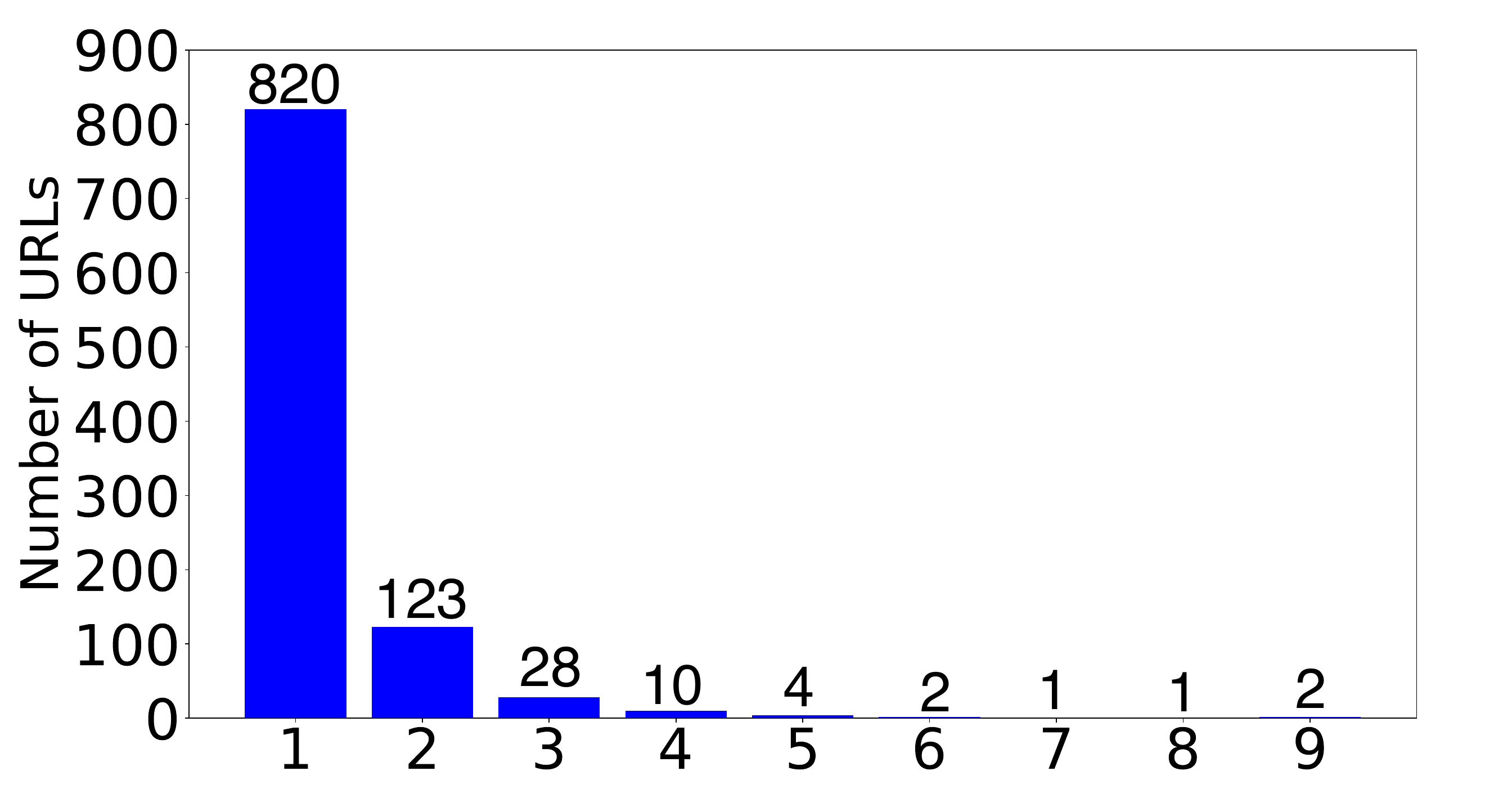}\label{fig:fig3}}
        \hfill
        \subfloat[False claim rate per user]{ \includegraphics[width=0.49\columnwidth]{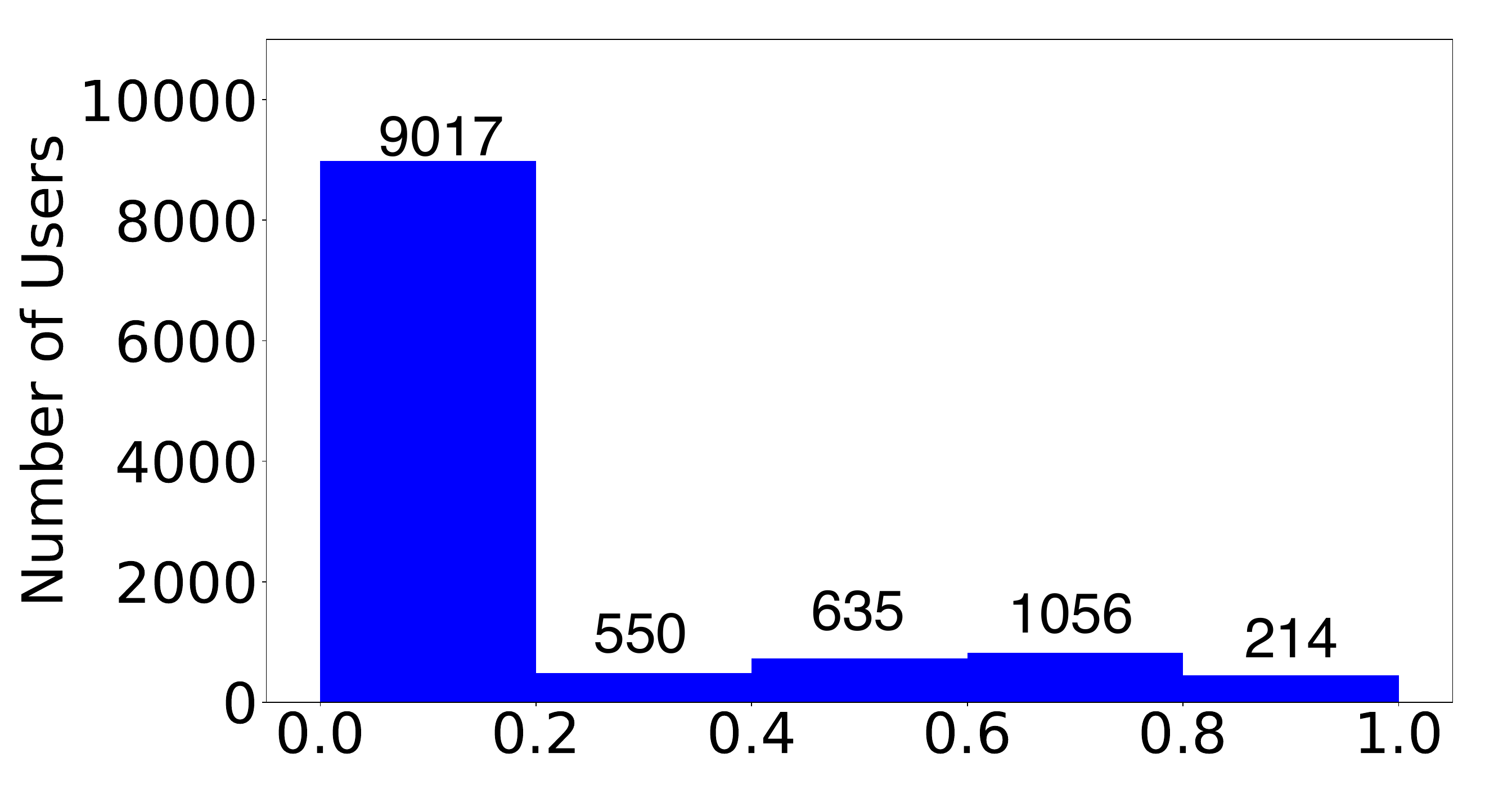}\label{fig:false-total-claims}}
         \caption{Histogram of tweets and users for false claims.
        }
  \label{fig:accounts}
\end{figure}

\subsubsection{Campaign Detection: Campaigns against Specific  Websites.} 

Figure~\ref{fig:fig3} shows the histogram of tweet counts for all the websites in our false claim dataset. While most of the websites have been falsely reported in only one tweet, there are some campaigns where a phishing claim against a specific website has been tweeted multiple times. 
We identified 48 URLs, that were tweeted more than 3 times, by a total of 32 unique users. This suggests the existence of  campaigns against specific websites. 

\subsubsection{Campaign Detection: Users with Many False Claims} 
Figure~\ref{fig:false-total-claims} shows the histogram of users with false claims divided by their total number of phishing claims. We found that only a small number of users have posted many phishing claims.
Interestingly, we found one user with 650 false claims and 3K true claims. 
About 78\% of the users in our dataset, i.e., 8,942 users, have only true claims, while 29 users have only \emph{false claims}. 
Also, 502 (4.37\%) users have a false claim rate of around 0.5. Almost all of these users have an equal amount of tweets, i.e., one true claim and one false claim (310 users) or two true claims and two false claims (174 users). 
Users that have only posted \emph{false claims} are suspicious and they might have deliberately sent these tweets.

\begin{figure}[t]
    \centering
    \includegraphics[width=0.75\columnwidth]{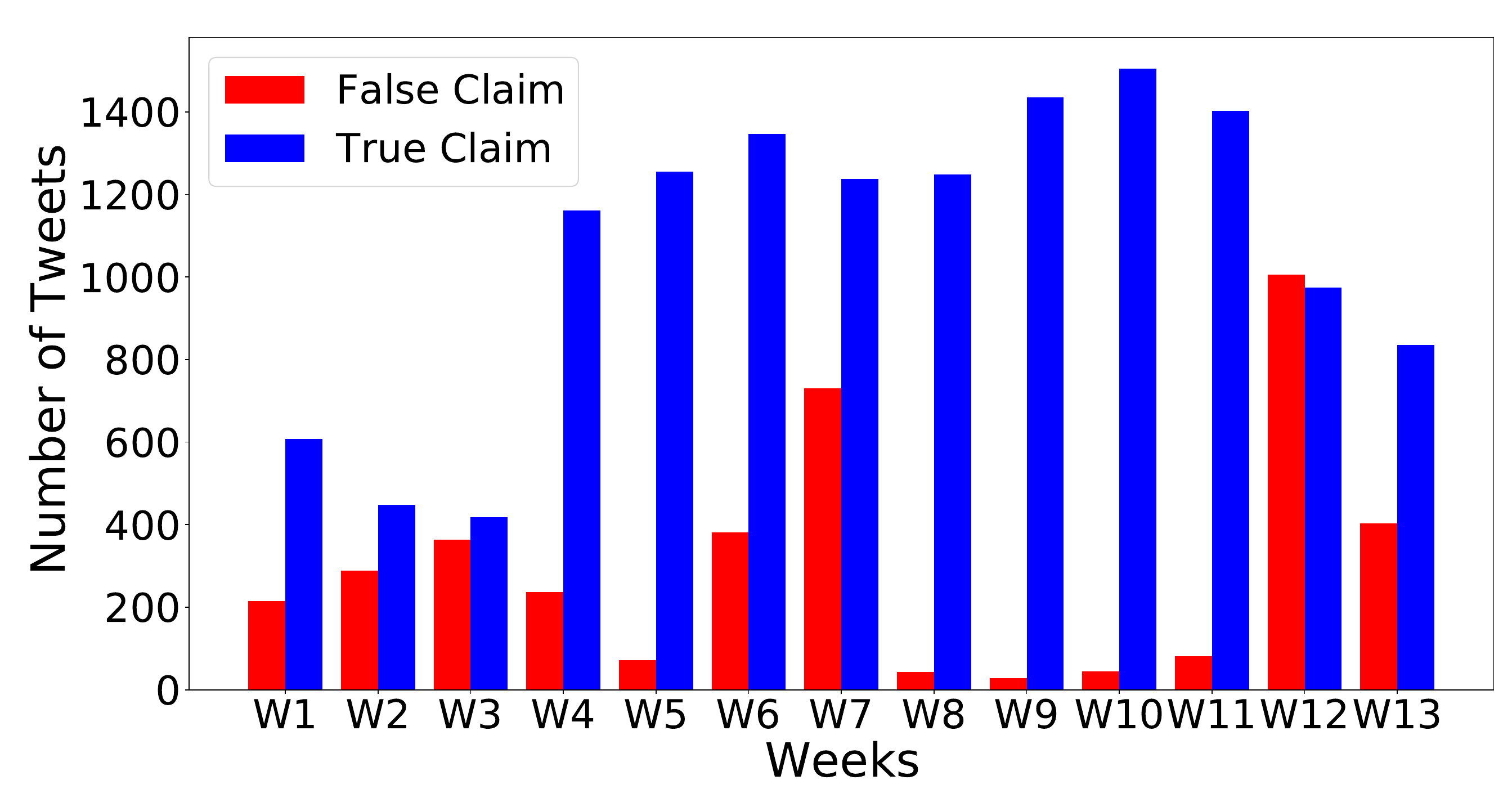}
    \caption{False vs. true phishing claims growth
    }
    \label{fig:phismisinformation-growth}
\end{figure}
\textbf{Bots.} We analyzed the false claim users using Botometer~\cite{sayyadiharikandeh2020detection}, a tool that gives a score to the accounts based on their social activity and other features. Using a threshold of 0.6, we found 2 bots and 139 real accounts, while 7 users changed their profiles to protect. 

\textbf{Growth of phishing reports.}
Figure~\ref{fig:phismisinformation-growth} shows the number of true and false phishing reports weekly from January 11, 2021 (the start of our data collection), to week~13 being April 11, 2021 (the end of our data collection). As you can see, false reports do not follow a trend, and in some weeks, there is a huge growth in their number, e.g., week~7 and week~12, while in some other weeks there is a huge drop in their number, e.g., in week~5 or week~8. 
In all weeks, however, except week~12, the number of true phishing reports is higher than that of the false reports.  The sharp increase in the number of false reports in week~12 is in-line with the real-world event when the COVID-19 Delta variant became the dominant variant and the number of phishing scams related to COVID-19 increased~\cite{delta-timeline}. 

\textbf{Discussion.} Our results show that not all phishing reports are reliable and there is a need for mechanisms to validate their correctness. Our proposed framework in  Figure~\ref{fig:phishing-datasetcollect} can be employed by social media platforms to detect and remove false phishing reports. 

\section{Framework for Detecting Misinformation about Zoom's Security and Privacy Threats}
\label{Zoom} 
\begin{figure*}[t]
    \centering
    
    \includegraphics[width=0.75\textwidth]{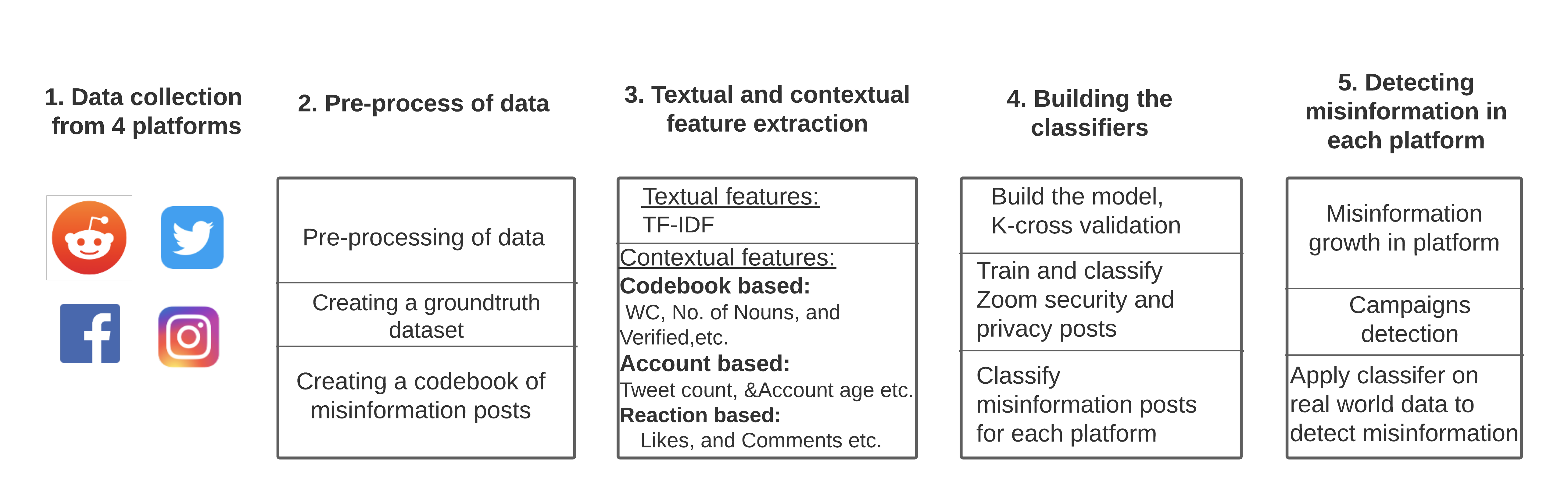}
    \caption{The framework developed for detecting Zoom security \& privacy misinformation.}
    \label{fig:zoom-datacollect}
\end{figure*}

In this section, we focus on social media discussions around Zoom's security and privacy threats, and propose a framework for detecting misinformation regarding it on Facebook, Instagram, Twitter, and Reddit. 
Analyzing public data from multiple social media platforms can also help us to investigate how misinformation is circulated differently on these platforms. 
To detect misinformation posts in each social media platform, we developed a binary classifier specific to that platform. Figure~\ref{fig:zoom-datacollect} shows our proposed framework: (1) Data collection, (2) Groundtruth and codebook creation, (3) Feature selection, (4) Training and testing  classifiers, and (5) Detecting the misinformation in each platform.

\begin{table}[ht]
\centering
\caption{The number of posts in each platform }
\label{table:initialdata}
\centering
    \resizebox{0.8\columnwidth}{!}{%
    \begin{tabular}{l| l  l  l}
    \hline 
    Platform & 2019 & 2020 & After filtering\\
    \hline 
    Instagram & 42,639 & 422,874 & 6,885 \\
    Facebook & 167,718 & 4,537,280 & 74,590 \\
    Reddit & 21,250 & 134,866 & 9,167\\
    Twitter & 45,178 & 1,011,022 & 870,852\\  
    \hline 
\end{tabular}
}
\end{table}

\begin{table*}[h]
\centering
\caption{Examples of posts and assigned labels.}
\resizebox{0.8\textwidth}{!}{%
\label{psotdec}
\begin{tabular}{c|p{0.6\textwidth}|p{0.15\textwidth}|p{0.25\textwidth}}
\hline \\[-1.8ex]
\multicolumn{1}{c|}{No.} & \multicolumn{1}{c|}{Post} & \multicolumn{1}{c|}{Label} & Reason  \\ \hline \hline\\[-1.8ex] 
1 & Security researchers have called Zoom ``a privacy disaster'' and ``fundamentally corrupt'' as allegations of the company mishandling user data snowball \#Data \#Breach \#Zoom  https://t.co/r3NcjsmuAB & Zoom's security \& privacy & Satisfies all the criteria and link goes to ``Guardian'' website   \\[+1.1ex] 
2 & @XXX CEO @XXX: With the popularity of \#Zoom, some security concerns have come to light-No end-to-end encryption for call- Sale of user data and analytics without disclosing or proper authorization. \#UpskillGang \#MilimaCyberAwareness @XXX' [Tweeted on April 24th 2020] & Misinformation & Fails to provide evidence and that no end-to end encryption is false (\cite{zoomwir}) and that it sells user data (\cite{Zoomblog})  \\ [+1.1ex] 
3 & [\#coronavirus] Japanese fashion brand now sells T-shirts for \#Zoom mtg. Change the color and design but basically only simple green T-shirts. Using the technology of virtual back ground and change as you like See below news. Seems to be nice! https://t.co/8lTMxtzKZb & Irrelevant  & Not about cybersecurity \\
\hline
\end{tabular}
}
\end{table*}
\begin{table}[h]
\caption{The size of groundtruth datasets per platform.
}
\centering
\resizebox{0.8\columnwidth}{!}{%
\begin{tabular}{c| ccc}
\hline 
Platform & Zoom security & Misinformation & Irrelevant\\
 \hline 
Instagram & 545 & 15 &  2,740 \\
Facebook & 560 & 42 & 2,734 \\
Reddit & 1,045 & 16 &  2,234\\
Twitter & 1,865 & 36 & 1,468  \\  
 \hline
\end{tabular}
}

\label{table:datase}
\end{table}

\subsection{Data Collection}
In order to collect data from Facebook, Instagram, Reddit, and Twitter, we used the ``posts/search'' endpoint of the CrowTangle API~\cite{crowdtangle}. The CrowdTangle API provides about 2\% of all public Facebook groups and pages, 2M+ public Instagram pages, and about 20K+ of most active sub-reddits~\cite{crowdtangletrack}. We also collected Twitter data using the Twitter V2 Archive Search endpoint~\cite{Twitter:2021}, which allows us to search access public Tweets from the complete archive dating back to the first Tweet in March 2006. 
We collected English posts sent from June 1, 2019, to Nov. 30, 2020, to examine if users were discussing security and privacy issues of Zoom before the pandemic, and how the discussions changed when the pandemic started. 
We initially obtained posts that include the keyword \emph{Zoom}. We considered retweets in our dataset.
 
\textbf{Pre-processing and filtering:} Table~\ref{table:initialdata} shows the data collected from the four platforms. Since we collected the posts that included the keyword ``Zoom,'' our dataset contained many posts not talking about security and privacy. To find additional keywords, we used the snowball sampling technique~\cite{goodman1961snowball}. We started by using a couple of seed keywords, including Zoom, Security, and Privacy. We then extracted posts from our dataset for the month of March for each respective platform. Using the seed keywords, we iteratively identified potential keywords that frequently co-occur with the seeds, adding them to our seed list only after manually ensuring they are closely related to our topic. After saturation was reached, we manually combined keywords into composites. 
In total, we identified 18 such phrases, all starting with word \emph{Zoom} and then the following words: \emph{Malware, Phishing, Virus, Security, Exploit, Hijacking, Bug, Hackers, Privacy, Backdoor, Hacked, Security Bug, Windows, Passwords, Windows Steal, Zoombombing, and Data}. 
We then used our new expanded keyword list, to filter out the posts. The last column of Table~\ref{table:initialdata} shows the final dataset that was obtained after our filtering.

\subsection{Groundtruth Dataset}
\label{groundtruth}

\subsubsection{Groundtruth Creation}
\label{groundtruth}
For training the classifiers, we first manually labeled a subset of the posts on each platform to create a groundtruth dataset. 
Creating a groundtruth is not a trivial task because we need to verify the correctness of claims and discussions. 
We used the following three criteria to label the posts: (a) The post is talking about ``Zoom,'' (b) The post is talking about Zoom's security or privacy, and (c) The post is either providing some evidence, i.e., links, videos, etc., from reputable blogs or articles that are verifiable, or not providing supporting evidence, but we could verify the claims by cross-checking them with the reputable sources. 
For that, we checked the post context and ran a Google search to determine if the post context is already addressed by the company or reputable sources and if the claim can be validated. 
Using these criteria, we defined three labels:
\emph{(1)~Zoom's security and privacy:} if a post satisfies all of the above mentioned criteria, 
\emph{(2)~Misinformation:} if the third criteria is not satisfied, and 
\emph{(3)~Irrelevant:} if it fails to satisfy either first or second criteria. 
Some examples of the posts that were labeled are shown in Table~\ref{psotdec}. 

\subsubsection{Annotation Process} 
To hand-label the posts, two authors annotated 13,200 posts (3,300 randomly chosen posts from each of Twitter, Instagram, Facebook \& Reddit). For inconsistent results, coders discussed how to resolve disagreements. To assess the inter-coder reliability, we performed a Cohen-Kappa test~\cite{schuster2004note}. The inter-rater agreement measured with Kappa score was 0.972, which shows almost perfect agreement.  
Table~\ref{table:datase} shows the groundtruth dataset that was obtained after the annotation. We can observe that there are a significant number of \emph{irrelevant} posts in our dataset, however, we do find evidence of misinformation in our dataset. We found 23 instances of users inviting other users to ``Zoombomb'' their classes or meetings. We found that 20 of these targets were for the platform \emph{Zoom}, while the remaining three were for \emph{Google Meet}.

\subsection{Qualitative Analysis of Misinformation Posts} We qualitatively analyzed the misinformation posts to understand their types, targets, and properties. These properties then can be used as features for classification. We created a hierarchical codebook of misinformation about Zoom's security and privacy threats, applying the open coding process~\cite{glaser2017discovery}. 
Following this process, one of the authors coded the misinformation posts identified in the previous subsection until no new categories emerged. To improve the quality of the categories, we used an iterative process~\cite{corbin1990grounded} so that new categories were added or existing ones were reorganized. To create the codebook, we followed certain guidelines: (1) Read through the posts, and identify themes and sub-themes; (2) While creating the categories, identify the motive and meaning of the post; and (3) Consider various features that can help in the identification of categories.

\begin{table}[h]
    \centering
    \caption{Description of categories identified by qualitatively analyzing the misinformation posts}
    \resizebox{0.9\columnwidth}{!}{%
    \begin{tabular}{l|c|p{0.4\textwidth}}
        \hline
        Class& Sub-Class & Description  \\ \hline
        Sources & - &Providing altered videos or photos that are not in the context to create confusion; posting URLs that are invalid or are redirected to a third party site. \\ [+1.1ex] 
        Structural & - &Post has all CAPS headline and content, and misspelling in the content.\\ [+1.1ex] 
        Network & - &Has a large audience, is verified by the platform. Two or more sources show the same news with the same context over time. \\[+1.1ex] 
        \hline
        Post Type& Accusation &Accusing various countries and/or businesses of wrongdoing without relevant evidence. \\[+1.1ex] 
        &Misleading &Misleading the audience, promoting and solidifying  a myth that rejects accepted narrative, is aligned towards one way of thinking and draws conclusions based on a limited number of facts. \\ [+1.1ex] 
        & Security&Post is about fake Zoombombing attacks, sponsors notion that using Zoom leads to hacking, data theft, leads to the backdoor for NSA, is malware and suggests no encryption used for chats and using Zoom leads to phishing attacks. \\ [+1.1ex] 
        &Privacy &Post suggests that users are being watched by government, promotes that user data is being mined by other companies, and sponsors the notion that since no encryption, anyone can read your chats.\\
        \hline
    \end{tabular}
    
    \label{tab:taxdescrip}}
\end{table}

\begin{figure*}[!htb]
    \centering
    \begin{minipage}{.55\textwidth}
        \centering
        \includegraphics[width=\textwidth]{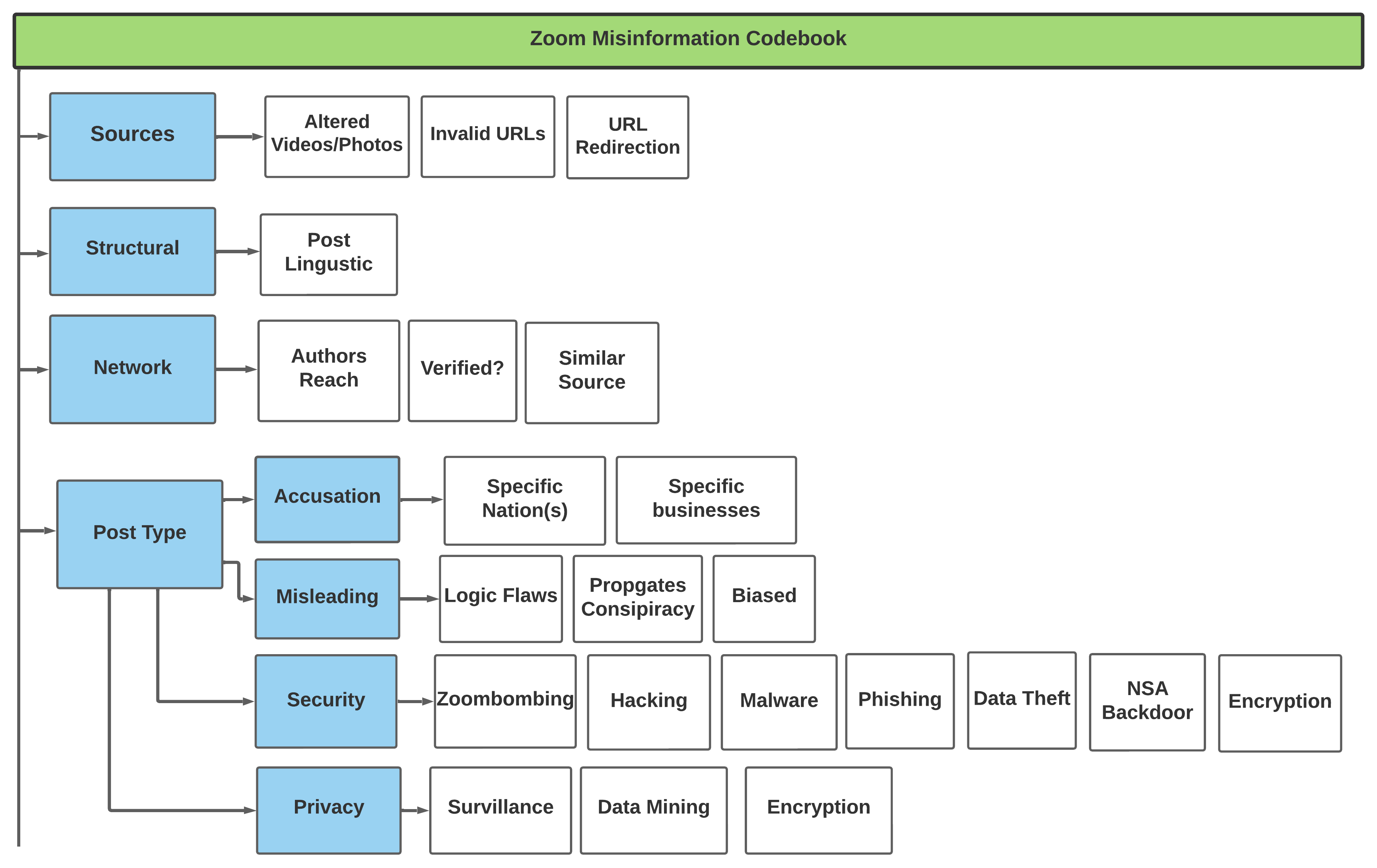}
        \caption{Zoom misinformation hierarchical codebook}
        \label{fig:taxonomy}
    \end{minipage}%
    \begin{minipage}{0.45\textwidth}
        \centering
            \captionof{table}{Performance of classifiers detecting posts about Zoom security and privacy. }
             \label{table:mlresults1}
            \resizebox{0.8\columnwidth}{!}{%
            \begin{tabular}{c| cccc}
            \hline 
            Platform & Accuracy & F1-Score & Precision & Recall\\
             \hline 
            Instagram  &  0.94  & 0.93  & 0.93  & 0.94  \\
             &  (+/- 0.02) &  (+/- 0.02) & (+/- 0.04) &  (+/- 0.01) \\
            Facebook  & 0.91 & 0.91  & 0.91 & 0.91  \\
               &  (+/- 0.02) & (+/- 0.02) & (+/- 0.02) &  (+/- 0.04) \\
            Reddit  & 0.93  &  0.93 & 0.93  &  0.93\\
               & (+/- 0.01)  & (+/- 0.02)  &  (+/- 0.02) & (+/- 0.03)  \\
            Twitter  & 0.81  & 0.81  & 0.81 & 0.81  \\ 
               & (+/- 0.02) &  (+/- 0.01) &(+/- 0.03) & (+/- 0.01) \\ 
             \hline
            \end{tabular}
            }
           \vspace{0.3cm}
            \captionof{table}{Percentage of Zoom's security and privacy posts.}
                \label{table:predictlabel}
                \resizebox{0.75\columnwidth}{!}{%
                \begin{tabular}{l| c  c }
                \hline 
                Platform & Zoom's security \& privacy & Irrelevant\\
                \hline 
                Instagram & 551 (8\%) & 3,034  \\
                Facebook & 11,400 (15\%) & 59,890\\
                Reddit & 627 (7\%) & 5,240 \\
                Twitter & 505,418 (58\%) & 365,434\\  
                \hline 
                \end{tabular}
                }
    \end{minipage}
\end{figure*}

Figure~\ref{fig:taxonomy} shows the hierarchical structure of the 
codebook. We discovered 4 main topics: (i) \emph{Sources}: Posts that provide misleading sources videos, URLs, or invalid links to other sites, (ii) \emph{Structural}: Posts containing irregularities in the content like misspellings or written in capitalization, (iii) \emph{Network}: Relates to the reach or perceived audience of the author, and (iv) \emph{Post Type}: subdivided into 4 categories talking about security or privacy and text that can be misleading or accusing and contain logic flaws, biased authors, or propagating conspiracy. 
Table~\ref{tab:taxdescrip} gives a high level overview of the description of each of these classes. 
These classes describe the data, and a post might fit multiple classes. For example, a post can be misleading suggesting that Zoom has no encryption and simultaneously accusing the company to be a spyware from the Chinese Communist Party.

After saturation, two authors coded the 109 posts labeled as misinformation. To find the agreement score, we gave a value of 1, if two had a perfect agreement, otherwise, we divided the number of labels by the number of possible values. Using this methodology, we found a substantial agreement of 72.3\%. The distribution of posts among different classes are: 
Sources (12), Structural (20), Network (11), Accusation (45), Misleading (20), Security (102), and Privacy (67). Note that one post could be assigned to multiple classes. 
 
\subsection{Detecting Zoom's Security and Privacy Threats} Initially, we tried to use a binary classifier to differentiate misinformation from all other posts. However, this yielded mediocre results. 
We then employed a two-step approach, where we first built a binary classifier to detect posts related to Zoom's security and privacy, and then built another classifier to detect misinformation among them. 
For the first classifier, we used the groundtruth dataset, where we used the misinformation posts and the Zoom's security and privacy posts. We built one classifier for each of the four platforms as each platform gives a different style of data, e.g., Twitter allows up to 280 characters while there is no constraint on that of Facebook posts. 
To build our supervised classifiers, we found that a mix of uni-gram \& bi-gram features provide a better result compared to uni-gram, bi-gram or tri-gram. Before extracting the features, we performed pre-processing on our dataset, removed stop words, emojis, special character (e.g., hashtags), and URLs. We vectorized our data using the TF-IDF. 
Since our groundtruth dataset for each platform was unbalanced (see Table~\ref{table:datase}), we employed several oversampling techniques.  We also tested multiple classification algorithms. To evaluate our classifiers, we used k-cross validation, where $k=3$.  
From our analysis we found that RandomOversampler was the best for Instagram, however, SMOTE provides better results for Facebook, Twitter and Reddit. Table~\ref{table:mlresults1} shows the classification performance of Random Forrest classifier as it provided the best accuracy across all four platforms. After developing classifiers detecting posts related to Zoom security and privacy, we ran the classifier on the whole datasets of all the platforms. 
Table~\ref{table:predictlabel} shows the percentages of security and privacy-related posts in our Instagram, Facebook, Reddit, and Twitter datasets, which are 8\%, 15\%, 7\%, and 58\%, respectively. 

\subsection{Detecting Misinformation about Zoom's Security and Privacy Threats}  
We trained another binary classifier for each platform to detect misinformation among posts relevant to Zoom security. 

\subsubsection{Feature Selection}
We used a combination of textual and contextual features. 
\textbf{Textual Features:} 
For each platform, we extracted bi-grams and uni-grams from all the posts and considered the top 100 of them with the highest values of TF-IDF, which resulted in 37 uni-grams and 63 bi-grams. We tested our classifiers with 100, 500, 1000, 1500, and 2000 top n-grams. While 500 provided the best results for Instagram, 100 provided the best results for other platforms. 
\textbf{Contextual Features:} We extracted a set of contextual features from the meta-data, including (1) 
\emph{Features inspired by the qualitative analysis:} Creating the codebook, we found some features being more apparent in misinformation. For example, in terms of network structure, they tend to have a large audience and are verified, or in terms of sources, they tend to provide altered videos or photos. 
Therefore, we used the following features: \emph{word counts}, \emph{noun counts}, \emph{pronouns counts}, the total number of capital words in a tweet, i.e., \emph{CAPS}, \emph{misspelled words count}, \emph{verified account}, \emph{followers count}, \emph{has a photo/video} and \emph{has a URL}. 
(2) \emph{Reaction-inspired features:} Posts can get reactions, such as \emph{likes}, \emph{retweets/shares}, \emph{comments}, etc. 
We used the following features: for Instagram, \emph{likes count}, for Facebook, the number of \emph{likes, comments}, and \emph{shares}, for Reddit, the number of \emph{likes}, and  \emph{comments}. For Twitter, we did not find the distribution of a number of \emph{likes} and \emph{retweets} being statistically different among the two classes, therefore we did not use them.
(3) \emph{Features based on account characteristics} 
including: \emph{tweets count}, \emph{profile description length}, \emph{account age}, \emph{listed count}, and \emph{has a profile image}. 

\subsubsection{Classifiers} 
Testing several oversampling techniques, we found that RandomOverSampler provides the best results for Instagram and Reddit classifiers, and SMOTE provides the best results for Facebook and Twitter classifiers. 
Also, we found that out of the five machine learning algorithms, Random Forest provides the best accuracy across the four platforms. We compared the results of the different algorithms after performing hyperparameter tuning, conducts an exhaustive search over the parameters to find the best combination of parameters. Table~\ref{table:completepredict} shows all classifiers, using k-cross validation ($k=3$), have great performances. 

 \begin{table}[h]
\centering
\caption{The performance of classifiers detecting misinformation about Zoom}
\label{table:completepredict}
\centering
    \resizebox{0.8\columnwidth}{!}{%
    \begin{tabular}{l| cccc}
    \hline 
    Platform & Accuracy & F1 Score & Precision & Recall\\
    \hline 
    Instagram & 0.98 &0.98 & 0.98   & 0.98 \\
       &  (+/- 0.01) & (+/- 0.02) &   (+/- 0.01) &  (+/- 0.02)\\
    Facebook & 0.99 &0.99  & 0.99  & 0.99  \\
      & (+/- 0.00)  & (+/- 0.01)  & (+/- 0.00)  &  (+/- 0.01) \\
    Reddit  & 0.99  & 0.99  & 0.99 & 0.99 \\
       & (+/- 0.01)  & (+/- 0.01)  & (+/- 0.00)  & (+/- 0.01) \\
    Twitter  & 0.98 & 0.98 & 0.98 &  0.98  \\ 
        & (+/- 0.01) &  (+/- 0.02) & (+/- 0.02) &  (+/- 0.01) \\ 
    
    \hline 
\end{tabular}
}
\end{table}

We examined feature importance in the trained Random Forest model to understand which of the features have a higher importance in the classification tasks. The top 5 features and their scores for each model are: 
\textbf{Facebook:} no. of all CAPS words (0.083), word count (0.081), \emph{company} (0.052), \emph{behind} (0.046), \emph{hey} (0.045). 
\textbf{Reddit:} \emph{company} (0.116), \emph{China} (0.087), no. of likes (0.066), word count (0.060), \emph{security} (0.050). 
\textbf{Instagram:} \emph{Zoom} (0.084), No. of all CAPS words (0.034), word count (0.033), \emph{security} (0.029), \emph{away} (0.029). 
\textbf{Twitter:} Has photo/video (0.072), URL in Tweet (0.070), account age (0.049), \emph{say} (0.043), has a profile image (0.042). 
The top features for each platform span a variety of feature categories, including textual features such as n-grams, and contextual features, consisting of reaction-, and account- inspired features. 

\subsubsection{Prevalence of misinformation about Zoom} 
Finally, we employed our trained classifiers on the posts that are related to security and privacy to detect those that are misinformation. 
We found that overall about 3\%, 18\%, 4\% and 3\% of posts on Instagram, Facebook, Reddit, and Twitter are \emph{misinformation}, respectively. 
We verified if our classifiers show consistent performance by manually labeling 200 (100 misinformation, 100 Zoom S\&P) posts on Twitter, 200 (100 misinformation, 100 Zoom S\&P) posts on Facebook, 116 (16 misinformation, 100 Zoom S\&P) posts on Instagram, and 125 (25 misinformation, 100 Zoom S\&P) from Reddit, and computing the accuracy and F1-score on these testing tests. Two authors coded the random set, and for disagreements, they discussed having a final label. The Cohen-Kappa score was 0.784, showing substantial agreement. We found that about 93\% and 92\% accuracy and F1-score in the case of Instagram, 94\% and 92\% in the case of Reddit, 93\%, and 93\% for Facebook, and 90\% and 90\% for Twitter. 

\begin{table}
\caption{Descriptive statistics of our final Twitter dataset.}
\centering 
\resizebox{\columnwidth}{!}{%
\begin{tabular}{lcccc|cccc}
\hline \\[-1.8ex] 
& \multicolumn{4}{c}{Users with true claims (N=220,159)} & \multicolumn{4}{c}{Users with misinformation (N=11,589)} \\
\\[-1.8ex] 
Feature & \multicolumn{1}{c}{Mean} & \multicolumn{1}{c}{Min} & \multicolumn{1}{c}{Max} & \multicolumn{1}{c}{Median}  & \multicolumn{1}{c}{Mean} & \multicolumn{1}{c}{Min} & \multicolumn{1}{c}{Max} & \multicolumn{1}{c}{Median} \\  
\hline \\[-1.8ex] 
Followers & 9,302 & 0 & $58.08M$ & 520& 10,451 & 0 & $8.4M$ & 675 \\ 
Friends & 1,582 & 0 & $1.06M$ & 678& 1,771 & 0 & $280K$ & 735\\
Tweets & 39,941 & 1 & $4.1M$ & 42,622& 10,911 & 1 & $3.04M$ & 10,534\\ 
Verified & 0.03 & 0 & 1 & NA & 0.03 & 0 & 1 & NA\\
\hline  

\end{tabular} 
}
 \label{stats-freq522} 
\end{table}

\subsection{Accounts and Campaign Characterization}
We studied characteristics of the accounts that posted misinformation, mostly focusing on the Twitter dataset, because CrowdTangle does not provide meta-data about the authors. 

\textbf{Descriptive Statistics:} Table~\ref{stats-freq522} statistically describes the account characteristics of users with true and false claims. 
The number of unique users with true and false claims was about 220K and 11K, respectively. If a user had posted both true and false claims, we considered them in both sets. 
To compare features, such as \emph{Followers, Friends, \& Tweets}, we ran Mann-Whitney U tests as they do not follow a normal distribution. 
We could reject the null hypothesis that users with \emph{false} and \emph{true} claims have the same distribution of followers count, and therefore, we can argue that on average, users with \emph{true} claims have lower number of followers than users with \emph{misinformation} claims ($Med_{true}=520$ vs. $Med_{mis}=675$, $p<0.0001$). We could not reject the null hypothesis for friends count \& tweet count. 
We ran chi-square test for the \emph{verified} variable, and found that accounts who post \emph{misinformation} are more likely to be verified ($M_{mis}= 0.03$ vs. $M_{true}=0.03$) ($X^2 =17.27 $, $p<0.0001$). This is interesting and also in-line with reports that \emph{verified} users are sharing misinformation at an all time high~\cite{Twitter-fake1,wang2018cure}.  

\subsubsection{Spread of Misinformation: Campaigns} 
We constructed a network based on \emph{following} relationships between 3,564 unique users in our misinformation dataset, in the month of April 2020, as it has the highest number of misinformation, using the Louvain Community Detection method~\cite{blondel2008fast}. 
In total, our network has 2,975 nodes and 22,934 edges. We could not collect the following list for 80 of the users because they made their profiles protected. We also found that 500 users were not connected to any of the users. The average weighted degree is 7.70. The average path length from one randomly selected node to another is 3.63. These values show that these nodes are very connected to each other. 
In total, we were able to obtain 35 communities, including some large ones with 671, 618, 585, 499, 194, 134, and 107 nodes. 
The biggest community has 22.55\% of all the nodes, while  13 communities have only 2 nodes. 

\begin{figure}[t]
    \centering
        \subfloat[CDF of \#retweets]{\includegraphics[width=0.49\columnwidth]{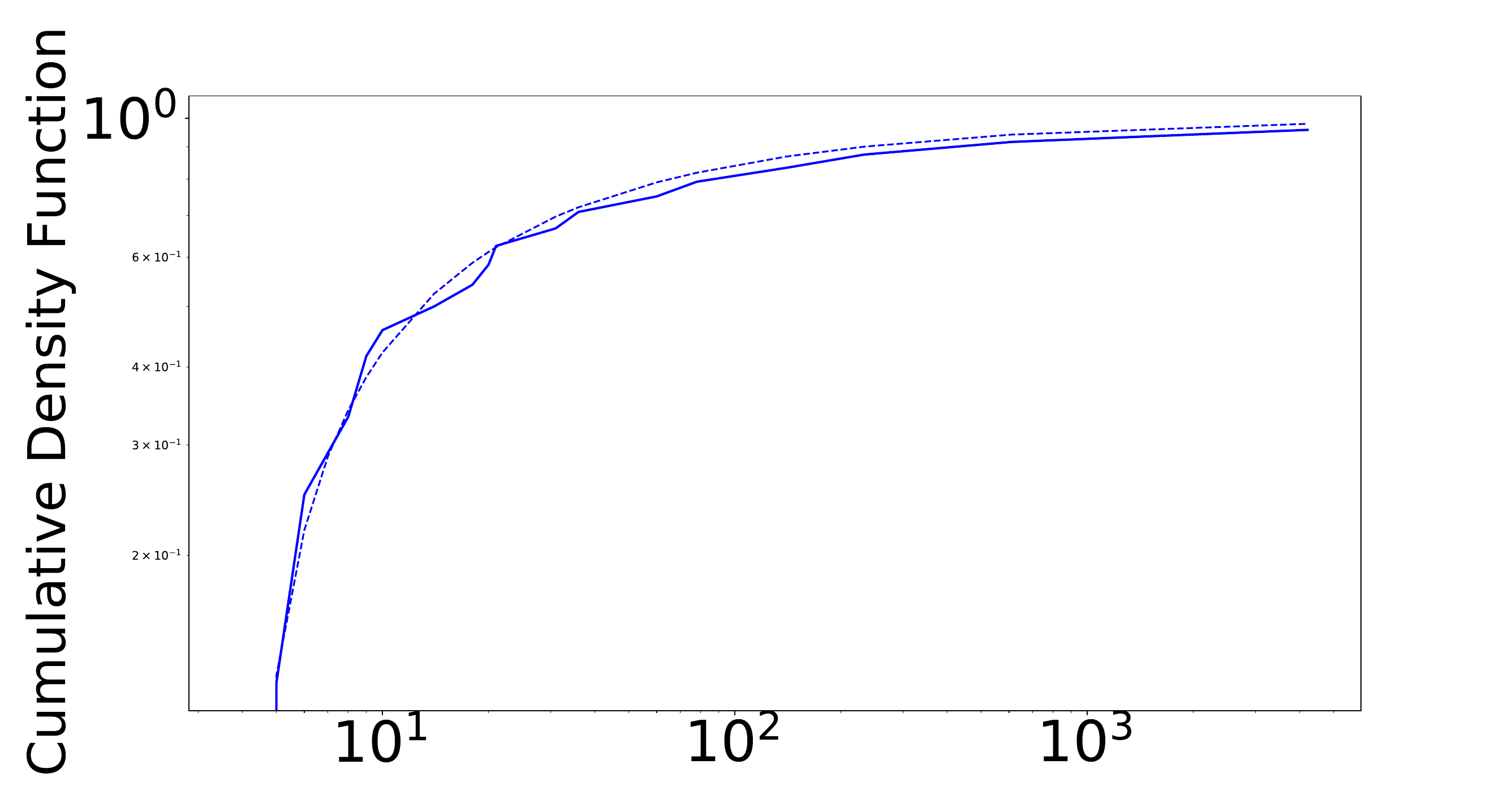}\label{fig:fig31}}
        \hfill
        \subfloat[Misinformation claim rate per user]{ \includegraphics[width=0.49\columnwidth]{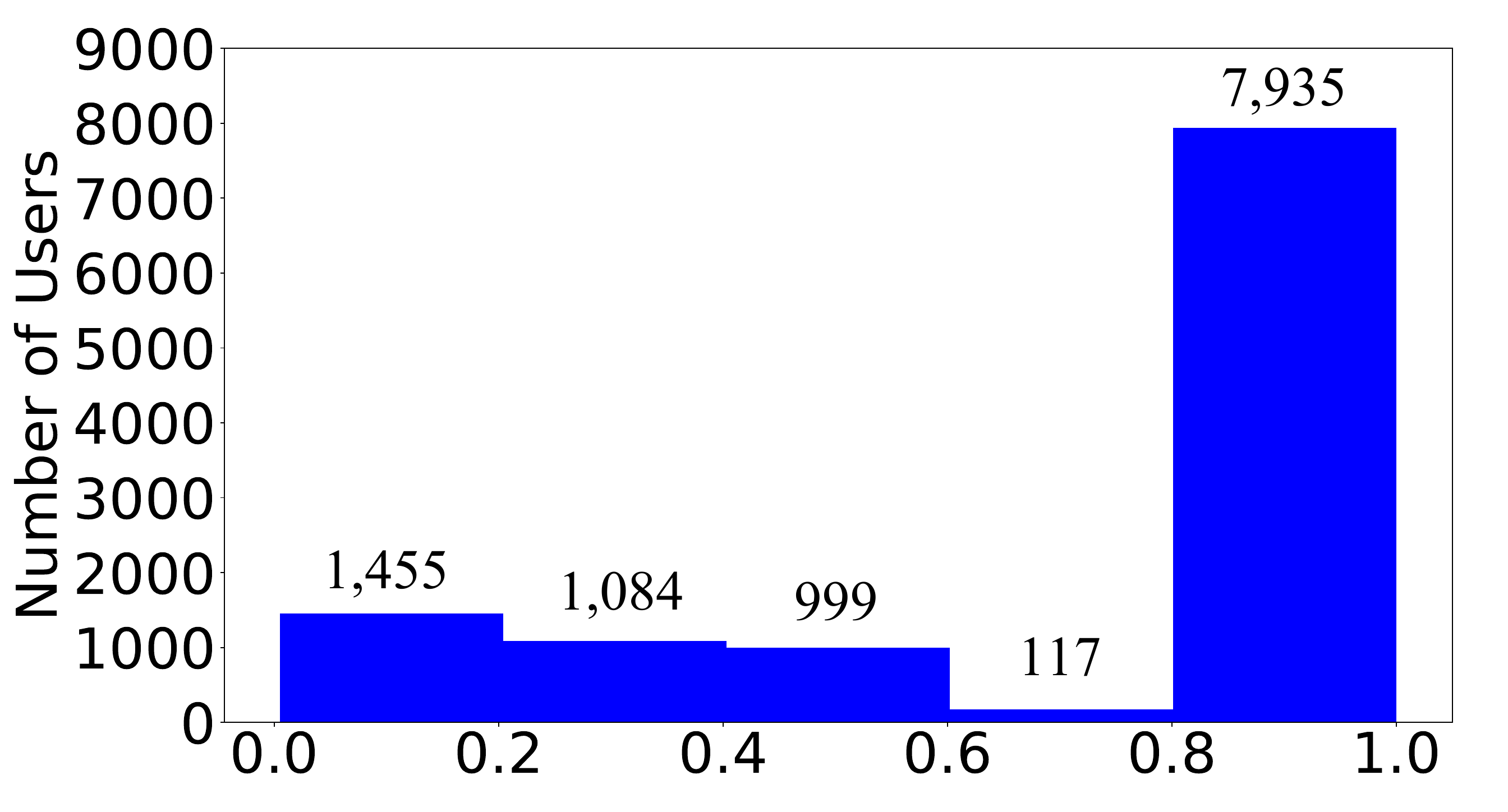}\label{fig:false-total-claims1}}
         \caption{CDF \& Histogram of tweets and users for misinformation claims.
        }
  \label{fig:accounts12}
\end{figure}

\subsubsection{Spread of Misinformation: Re-Tweet Count:} 
Figure~\ref{fig:fig31} shows the CDF of retweet counts for all the tweets in our misinformation dataset. While most of the misinformation was tweeted only once, we found some campaigns. We found one misinformation tweet that was retweeted 855 times. This retweet was sent by 854 unique users. The topic of the tweet was accusing Zoom of being a Chinese Communist Party malware and that they are using it for surveillance.

\subsubsection{Campaign Detection: Users with Many False Claims:} 
Figure~\ref{fig:false-total-claims1} shows the histogram of users with misinformation claims divided by their total number of Zoom claims. 
We found that a majority number of users have posted many Zoom misinformation claims.
Interestingly, we found one user with 10 misinformation claims and 2K Zoom S\&P claims. 
About 68\% of the users in our dataset, i.e., 7,935 users, have only \emph{misinformation claims}. 
Also, 955 (8.24\%) users have a false claim rate of around 0.5. Almost all of these users have an equal amount of tweets, i.e., one true claim and one false claim (842 users) or two true claims and two false claims (113 users). 

\textbf{Bots.} We analyzed all the misinformation users using Botometer~\cite{sayyadiharikandeh2020detection}, and threshold of 0.6. We found 208 bots and 11,274 real accounts, while 107 users had changed their profiles to be protected. 

\begin{figure}[t]
    \centering
    \includegraphics[width=\columnwidth]{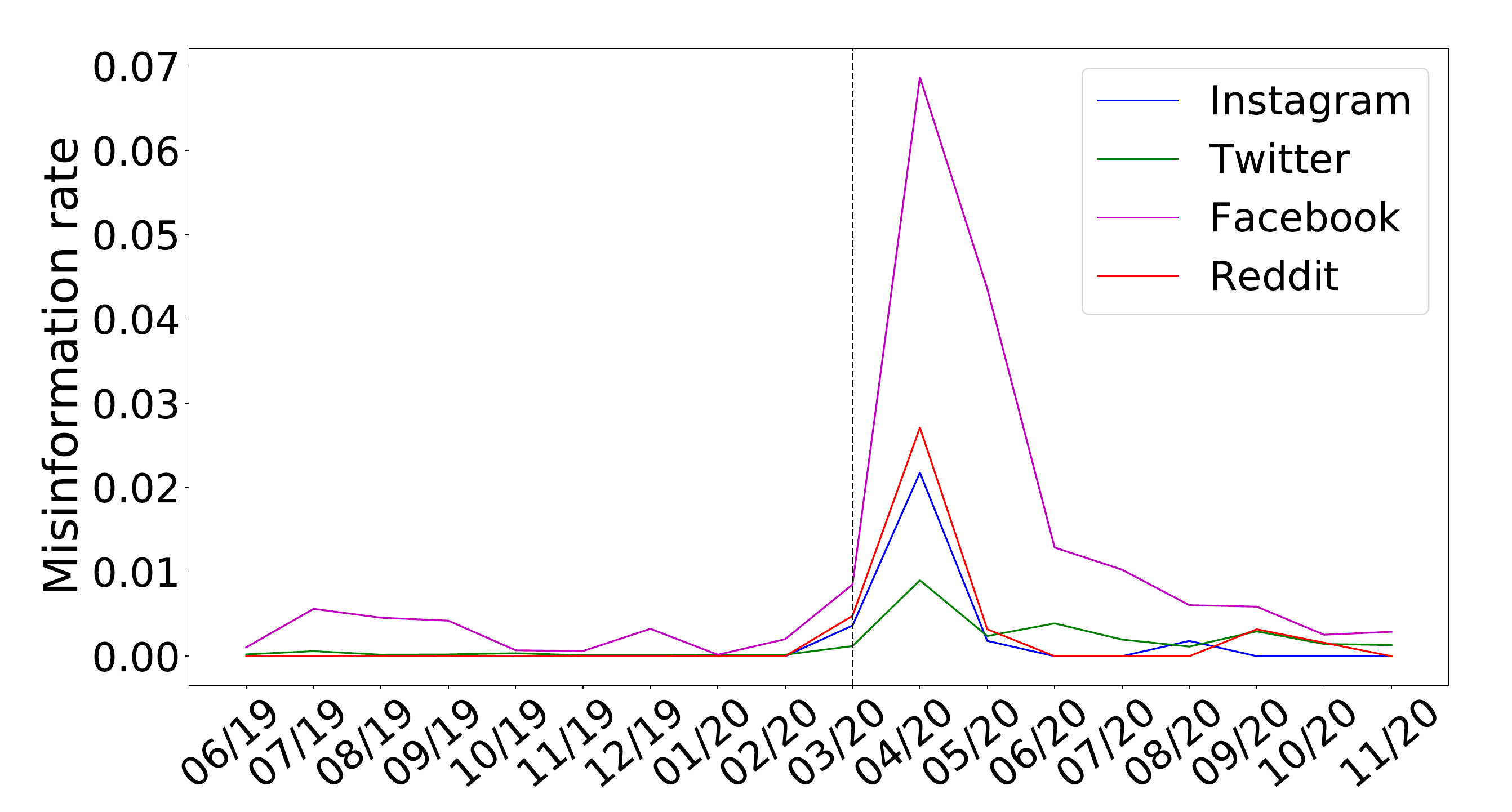}
    \caption{Misinformation growth rate.}
    \label{fig:misinformation-growth}
\end{figure}

\textbf{Growth of Misinformation}
Figure~\ref{fig:misinformation-growth} shows the percentage of misinformation over time on the four social media platforms. The vertical dotted black line represents the time when multiple states in the US went into COVID-19 lockdown~\cite{covid-timeline}. 
As you can see, on Facebook, there are some misinformation about Zoom since 2019. We manually inspected these posts and found that in the month of July 2019, users were discussing \emph{Zoom hit by DoS}. However, while Zoom indeed had a new zero-day vulnerability that could be used to employ a DoS attack, no actual attack happened~\cite{zoom-zero-day}. 
We also see a sudden spike, in the number of misinformation posts on Facebook around February 2020, and then subsequent spikes in Instagram, Twitter, and Reddit after March 2020. Our analysis revealed that users were claiming that Zoom is malware, a tool by the Chinese Communist party to spy on people, etc., however, these claims have been refuted~\cite{zoomwir, Zoomblog}. 
The plot shows a higher percentage of misinformation posted on Facebook when compared to other platforms, and at its peak, it gets to 7\% of posts being misinformation. 

\textbf{Discussion.} We showed that misinformation about Zoom security and privacy was spread on all the social media platforms. We argue that there is a need for mechanisms validating information about technological topics. 
Our proposed framework in  Figure~\ref{fig:zoom-datacollect} is the first step toward such a goal.

\section{Ethical Considerations and Broader Impact }

We analyzed publicly available data, provided by Twitter Streaming API and Crowdtangle API. We also follow standard ethical guidelines~\cite{rivers2014ethical}, not making any attempts to track users across sites or deanonymize them. We believe that our results show that misinformation about cybersecurity and privacy exists, and we hope that the community can further investigate its impact on the user and can research solutions to tackle this challenge.

\section{Limitations and Future Work}
The analysis on phishing misinformation gives a lower bound of the \emph{misinformation} on Twitter because we had access to a 1\% sample of Twitter data. 
Similarly, in our second study, the size of our datasets was restricted by \emph{CrowdTangle}. 
Also, not having access to the followers and friends of users on other social media platforms, we could only detect and analyze possible campaigns on Twitter. 
In the future, we would investigate the diffusion of cybersecurity and privacy misinformation and examine if they are different from other types of misinformation. 
We could also explore using semi-supervised techniques instead of supervised learning.

\section{Conclusion}
In this work, we proposed two frameworks for detecting misinformation about cybersecurity and privacy threats on social media, focusing on two topics with different types of misinformation: \emph{phishing websites} and \emph{Zoom’s security \& privacy threats}. 
We examined the correctness of Twitter reports posted by users about websites being phishing. In total, we found that about 9\% of all obfuscated URLs and about 22\% of tweets about phishing websites are \emph{misinformation}. 
Second, using a set of textual and contextual features, we built supervised classifiers to identify posts discussing the security and privacy of Zoom, and to detect misinformation in our whole dataset. Our classifiers showed great performance across all four platforms.
We found about 3\%, 18\%, 4\%, and 3\% of posts on Instagram, Facebook, Reddit, and Twitter, as misinformation, respectively. 
Our results show that misinformation about cybersecurity and privacy is present on social media, and the community needs to further study its impact on end-users and threat intelligence tools. 

\section{Acknowledgment}
This work is partially supported by NSF under
CNS-1932574 and a Comcast Innovation Fund. We thank Dr.~Mainack Mondel for helpful comments.

\end{document}